\documentclass[
12pt, aps, floatfix, pra, reprint, superscriptaddress, numerical, longbibliography ]{revtex4-1}

\usepackage{amsmath,amssymb,bm}
\usepackage{graphicx}
\usepackage{rotating}
\usepackage{adjustbox}
\usepackage{subcaption}
\usepackage{float}
\usepackage{mathrsfs}
\usepackage{xcolor}
\usepackage{xurl}
\usepackage[hidelinks]{hyperref}
\usepackage{xr}
\usepackage[percent]{overpic}
\usepackage{upgreek}
\usepackage{ragged2e}
\usepackage[compatibility=false]{caption}

\newcommand{\ii}{\mathrm{i}}
\newcommand{\dd}{\mathrm{d}}
\newcommand{\ve}[1]{\bm{#1}}
\newcommand{\unit}[1]{\,\mathrm{#1}}
\begin{document}

\title{Stabilization of counter propagating currents by energy conservation in collisions}

\author{Thomas Kluge}
\email{t.kluge@hzdr.de}
\affiliation{Helmholtz-Zentrum Dresden-Rossendorf, Bautzner Landstra\ss e 400, 01328 Dresden, Germany}
\author{Jannis Schulz}
\affiliation{Helmholtz-Zentrum Dresden-Rossendorf, Bautzner Landstra\ss e 400, 01328 Dresden, Germany}
\affiliation{Dresden University of Technology, 01069 Dresden, Germany}
\author{Arthur Hirsch-Passicos}
\affiliation{Helmholtz-Zentrum Dresden-Rossendorf, Bautzner Landstra\ss e 400, 01328 Dresden, Germany}
\author{Pawe\l{} Ordyna}
\affiliation{Helmholtz-Zentrum Dresden-Rossendorf, Bautzner Landstra\ss e 400, 01328 Dresden, Germany}
\affiliation{Dresden University of Technology, 01069 Dresden, Germany}
\author{Ulrich Schramm}
\affiliation{Helmholtz-Zentrum Dresden-Rossendorf, Bautzner Landstra\ss e 400, 01328 Dresden, Germany}
\affiliation{Dresden University of Technology, 01069 Dresden, Germany}

\date{\today}

\begin{abstract}
Current filamentation generates magnetic fields and shapes electron transport in dense plasmas, where collisions can either suppress or enhance its growth.
Here we show that conserving electron kinetic energy in a collisional model can change both the dominant instability and its characteristic wavelength.
We develop a linear kinetic theory for drifting, anisotropic electron populations with a relaxation operator that conserves particle number and laboratory-frame kinetic energy while allowing momentum transfer to a stationary ion background.
For counterstreaming currents, energy conservation maintains the stabilizing influence of charge separation over a broader range of collision frequencies than a model conserving particle number alone.
At intermediate collision rates, the resulting suppression of transverse filamentation allows a nearly aperiodic oblique mode to dominate, reducing the maximum growth rate and shifting the wavelength of the fastest-growing perturbation.
Two-dimensional particle-in-cell simulations support predicted growth rates of up to more than a magnitude lower than expected without energy conserving collision.
\end{abstract}

\maketitle

Counterstreaming currents and temperature anisotropy drive electromagnetic instabilities that generate magnetic fields and reshape electron transport \cite{Weibel1959,Fried1959,Bret2004,Bret2005a,Bret2009,Bret2010a}.
The current filamentation instability (CFI) and related Weibel modes are ubiquitous and occur e.g. in fusion-relevant plasmas \cite{Okada1989,Hill2005,Robinson2008b,Ruyer2026}, relativistic beam transport \cite{Lee1973,Bret2010a}, dense laser-produced plasmas \cite{Cottrill2008,Fiore2010b,Metzkes2014,Gode2017,Scott2017,Ruyer2020b,Raj2020,Ordyna2024,Schoenwaelder2025}, accelerator experiments \cite{Allen2012,Shukla2018b}, and collisionless shocks, where magnetic-field generation contributes to shock formation and particle acceleration \cite{Medvedev1999b,Silva2003b,Bret2010a}.

Kinetic theories describe CFI across relativistic, finite-temperature regimes \cite{Bret2005a,Bret2010a,Allen2012,Hao2009a}.
When forward and return currents pinch unequally, charge separation produces an electrostatic restoring field that can suppress filamentation and shift the dominant wavevector \cite{Tzoufras2006b}.
Collisions further modify the mode competition in dense plasmas \cite{Honda2004,Karmakar2008,Hao2009b,Fiore2010b,Hajisharifi2017}, which is relevant e.g. for laser isochoric heating, compression or ion acceleration.
There, laser-generated electrons drive a compensating return current \cite{Glinsky1995,Huang2016,Kluge2018c,Yang2024,Huang2026,Bell1997a} and generate magnetic fields through filamentation \cite{Tabak1994,Bell1997a,Kodama2001,Gremillet2002,Sentoku2000b,Honda2000,Sentoku2003a,Honrubia2008,Metzkes2014}.
Depending on plasma conditions, collisions can damp filamentation or support a resistive branch \cite{Hao2009b,Hao2009a,Cottrill2008,Fiore2010b,Ruyer2020b}.
Enhancement is commonly attributed to suppression of the collective return-current response \cite{Molvig1975,Okada1980,Hao2008,Hao2009a}.
Collisions can also weaken the coupling to stabilizing space-charge fields \cite{Hao2014}.

Relaxation models of the Bhatnagar--Gross--Krook (BGK) type\cite{Bhatnagar1954} provide for an analytically tractable kinetic description of collisions.
Previous filamentation models have imposed none or only particle-number conservation while leaving electron kinetic energy unconstrained \cite{Fiore2010b,Robinson2014, Schoenwaelder2025}.
For nearly elastic scattering from heavy, stationary ions, however, electrons transfer momentum while approximately conserving their kinetic energy in the ion rest frame.
Energy-conserving relaxation theories exist for longitudinal response and more general plasma settings \cite{Fried1966,Opher2002,Formanek2021,Chuna2025}, motivating an examination of how this constraint affects electromagnetic mode competition in counterstreaming plasmas.
Here, we develop an analytic expression for the conductivity that imposes energy conservation, and find that in certain cases of broad-energy driver beams it can reduce instability growth rates by more than an order of magnitude.

We consider a homogeneous equilibrium plasma composed of one or more species \(s\), with a general three-dimensional momentum-space distribution \(f_0^{(s)}(\ve p)\) normalized to \( \int f_0^{(s)}(\ve p)\,\dd^3p=n_0^{(s)}\).
In the applications considered at the end of this Letter a non-relativistic drifting anisotropic Gaussian will be used exemplarily to derive closed-form expressions (see Supplemental Material~\cite{SupplementalMaterial}) and numerical results (Sec.~III).
We decompose the full distribution and fields as \(f^{(s)}=f_0^{(s)}+\delta f^{(s)}\) and \(\ve E=\delta\ve E\), \(\ve B=\delta\ve B\), with homogeneous equilibrium \(\ve{\nabla} f_0^{(s)}=0\), \(\ve E_0=\ve B_0=0\), and small perturbations \(\delta\) proportional to $ \exp\left[\ii(\ve k\cdot\ve x-\omega t)\right]$ with wavevector $\ve k=(k_x,k_y,k_z)$ and frequency $\omega=\omega_r+\ii\Gamma$.
The perturbed current is a three-dimensional linear function of the electric field, $\ve J=\bm{\upsigma}\ve E$, which defines the conductivity tensor \(\bm{\upsigma}\).
The corresponding dielectric tensor is \( \bm{\upvarepsilon} = \mathbf I + \ii\bm{\upsigma}/\omega \).

Starting from Maxwell's equations \( \ve{\nabla}\times\ve E=-\partial_t\ve B \), \( \ve{\nabla}\times\ve B=\ve J+\partial_t\ve E \), and Fourier transforming according to \(\ve{\nabla}\to\ii\ve k\) and \(\partial_t\to-\ii\omega\), one obtains \(\ve k\times\ve E=\omega\ve B\) and \( \ve k\times\ve B=-\omega\ve E-\ii\ve J \).
Eliminating \(\ve B\) gives \( (\omega^2-k^2)\ve E+\ve k(\ve k\cdot\ve E) = -\ii\omega\ve J\).
Using $\ve J=\bm{\upsigma}\ve E$, this becomes the wave equation
\begin{equation}
\mathbf W(\omega,\ve k)\ve E=0, \qquad \mathbf W = \bm{\upvarepsilon} - \frac{k^2}{\omega^2} \left( \mathbf I-\frac{{\ve k}\otimes{\ve k}}{k^2} \right),
\label{eq:M_def}
\end{equation}
which has non-trivial solutions where $\det\mathbf W(\omega,\ve k)=0$.
Perturbations are growing, i.e. the system is unstable, for solutions with $\Gamma > 0$.
Here and in the following we employ normalized units \(c\equiv e\equiv m_{e,0}\equiv1\) and normalize frequencies to the effective plasma frequency \(\omega_\mathrm{eff}=\sqrt{n_\mathrm{eff}}\), where $n_\mathrm{eff}=\sum_s n_0^{(s)}\left(q^{(s)}\right)^2/m^{(s)}$ and \(q^{(s)}\) and \(m^{(s)}\) denote the charge and mass of species \(s\), respectively.
For readability, the kinetic derivation below is written for one species with the species index suppressed.

To find solutions to the wave equation \eqref{eq:M_def}, we need to obtain $\mathbf \upsigma$, i.e. we are looking for an expression for $\ve J$ linearized in $\ve E$.
We start with linearized Vlasov-BGK equation
\begin{equation}
\frac{\partial \delta f}{\partial t} +\ve v(\ve p)\cdot\ve{\nabla}\delta f +q\left(\delta\ve E+\ve v(\ve p)\times\delta\ve B\right)\cdot\ve{\nabla}_{\ve p}f_0 = -\mathcal{C}[\delta f].
\end{equation}

We first construct the energy-conserving collision operator $\mathcal{C}$ by introducing the vector
\begin{equation}
\ve\psi(\ve p) =
\begin{pmatrix}
1\\
K(\ve p)
\end{pmatrix}
\label{eq:psi_def}
\end{equation}
of weights of the conserved moments $n=\int f(\ve p)\,\dd^3p$ and $\mathcal K=\int K(\ve p)f(\ve p)\,\dd^3p$, where
\begin{equation}
K(\ve p)
=
H(\ve p)-\langle H\rangle_0,
\qquad
\langle H\rangle_0
=
\frac{1}{n_0}
\int H(\ve p)f_0(\ve p)\,\dd^3p
\label{eq:K_main_def}
\end{equation}
is the centered single-particle kinetic energy.
Here, $H$ is the single-particle kinetic energy in the lab frame (i.e. ion rest frame).
Particle and energy conservation for the collisions then require
\begin{equation}
\int \ve\psi(\ve p)\, \mathcal{C}[\delta f]\, \dd^3p = 0.
\label{eq:moment_basis}
\end{equation}
We also define the perturbation moment vector
\begin{equation}
\delta\ve m
\equiv
\int \ve\psi(\ve p)\, \delta f(\ve p)\,\dd^3p
=
\begin{pmatrix}
\delta n\\
\delta\mathcal K
\end{pmatrix}.
\label{eq:nK_def}
\end{equation}
Here, $\delta\mathcal K$ denotes the perturbation of the centered kinetic-energy moment, with $\delta\mathcal K \equiv \int K(\ve p)\,\delta f(\ve p)\,\dd^3p$ by construction.
Finally, defining the equilibrium moment matrix
\begin{equation}
\mathbf S \equiv \frac{1}{n_0} \int \ve\psi(\ve p)\otimes\ve\psi(\ve p)\, f_0(\ve p)\,\dd^3p,
\label{eq:moment_gram_matrix}
\end{equation}
the minimal number- and energy-conserving BGK operator takes the form
\begin{equation}
\mathcal{C}[\delta f] = \nu \left[ \delta f - \frac{f_0}{n_0} \ve\psi^{\mathrm T} \mathbf S^{-1} \delta\ve m \right].
\label{eq:bgk_nK}
\end{equation}
The construction of this projector and the proof of simultaneous particle-number and laboratory-frame energy conservation are given in the Methods~\ref{app:conservation}.
Note that in the isotropic, non-drifting, longitudinal Gaussian limit, the operator reduces to the energy-conserving Krook operator introduced by Fried, Kaufman, and Sachs (FKS) \cite{Fried1966}.

The conductivity tensor now follows from the definition of the current $ \ve J = q\int \ve v(\ve p)\, \delta f(\ve p)\,\dd^3p $ after consistently factoring out $\ve E$.
We start by decomposing the perturbed distribution as \( \delta f=\delta f^{(E)}+\delta f^{(nK)} \), where \(\delta f^{(E)}\) is the direct Lorentz-driven response and \(\delta f^{(nK)}\) is the BGK-projector contribution.
Defining the resonance denominator $ D(\ve p) = \nu-\ii\left[ \omega-\ve k\cdot\ve v(\ve p) \right] $, the linearized Vlasov-BGK equation can then be written as
\begin{equation}
D\,\delta f = -q \left( \ve E+\ve v\times\ve B \right)\cdot \ve{\nabla}_{\ve p}f_0 + \frac{\nu f_0}{n_0} \ve\psi^{\mathrm T} \mathbf S^{-1} \delta\ve m.
\label{eq:delta_f}
\end{equation}
Thus
\begin{align}
\delta f^{(E)}& = -\frac{q}{D} \left( \ve E+\ve v\times\ve B \right)\cdot \ve{\nabla}_{\ve p}f_0,
\label{eq:delta_f_E}\\
\delta f^{(nK)} &= \frac{\nu f_0}{n_0D} \ve\psi^{\mathrm T} \mathbf S^{-1} \delta\ve m.
\label{eq:delta_f_nK}
\end{align}

Using Faraday's law \( \ve B=(\ve k\times\ve E)/\omega \), and defining the Lorentz-response tensor
\begin{equation}
\mathbf F(\ve p) = \left( 1-\frac{\ve k\cdot\ve v(\ve p)}{\omega} \right)\mathbf I + \frac{\ve k\otimes\ve v(\ve p)}{\omega},
\label{eq:F_def}
\end{equation}
so that $ \ve E + \ve v(\ve p)\times \left(\ve k\times\ve E\right)/\omega=\mathbf F(\ve p)\ve E, $ and using the dyadic product identity $\ve a\left(\ve b\cdot \ve c\right)=\left(\ve a \otimes \ve b\right)\ve c$, the direct field-driven current becomes
\begin{equation}
\begin{aligned}
\ve J^{(E)}
           & = -q^2 \left[ \int \frac{ \left( \ve p\otimes\ve{\nabla}_{\ve p}f_0 \right) \mathbf F(\ve p) }{ \gamma m\,D(\ve p) } \,\dd^3p \right] \ve E
\end{aligned}
\label{eq:J_E_conductivity}
\end{equation}
where the term in front of $\ve E$ readily defines the field driven part $\bm{\upsigma}^{(E)}$ of the conductivity.

For the BGK contribution to the conductivity, starting from Eq.~\eqref{eq:delta_f_nK} and using the symmetry of $\mathbf S^{-1}$ so that $\ve p \psi^T\mathbf S^{-1} =\ve p \left(\mathbf S^{-1}\ve \psi\right)^T=\ve p \otimes \left[ \mathbf S^{-1}\ve\psi \right]$, the BGK part of the current is found to be generated from the conserved-moment vector by the response matrix
\begin{equation}
\mathbf N \equiv \frac{q}{n_0} \int \ve p \otimes \left[ \mathbf S^{-1}\ve\psi(\ve p) \right] \frac{f_0(\ve p)}{\gamma m D(\ve p)} \,\dd^3p.
\label{eq:N_tensor}
\end{equation}
The BGK current is therefore simply
\begin{equation}
\ve J^{(nK)} = \nu\,\mathbf N\,\delta\ve m.
\label{eq:J_nK}
\end{equation}
It remains to determine the conserved-moment response \(\delta\ve m\).
Putting the decomposition \(\delta f=\delta f^{(E)}+\delta f^{(nK)}\) into the definition of $\delta \ve m$, Eqn.~\eqref{eq:nK_def}, gives $ \delta\ve m = \delta\ve m^{(E)} + \delta\ve m^{(nK)} $.
Using \eqref{eq:delta_f_E}, the field-generated moment source is
\begin{equation}
\delta\ve m^{(E)} \equiv \int \ve\psi(\ve p)\, \delta f^{(E)}(\ve p)\,\dd^3p = \mathbf G\,\ve E,
\label{eq:G_matrix}
\end{equation}
with
\begin{equation}
\mathbf G = -q \int \frac{ \left(\ve\psi(\ve p) \otimes \ve{\nabla}_{\ve p}f_0 \right) \mathbf F(\ve p) }{ D(\ve p) } \,\dd^3p
\label{eq:G_definition}
\end{equation}
The BGK-generated contribution follows from Eq.~\eqref{eq:nK_def} and \eqref{eq:delta_f_nK},
\begin{align}
\delta\ve m^{(nK)} & = \int \ve\psi(\ve p)\, \delta f^{(nK)}(\ve p)\,\dd^3p
                  = \nu \mathbf M \delta m
                  \label{eq:moment_nK}
\end{align}
with
\begin{equation}
\begin{aligned}
\mathbf M & \equiv \frac{1}{n_0} \int \ve\psi(\ve p) \otimes \left[ \mathbf S^{-1}\ve\psi(\ve p) \right] \frac{f_0(\ve p)}{D(\ve p)} \,\dd^3p
\end{aligned}
\label{eq:moment_integrals}
\end{equation}
so that Eq.~\eqref{eq:moment_nK} becomes
\begin{equation}
\delta\ve m^{(nK)} = \nu\,\mathbf M\,\delta\ve m.
\end{equation}
Using this result together with Eq.~\eqref{eq:G_matrix} gives
\begin{equation}
\delta\ve m = \left( \mathbf I-\nu\mathbf M \right)^{-1} \mathbf G\,\ve E.
\label{eq:moment_solution}
\end{equation}
Inserting this into Eq.~\eqref{eq:J_nK} and comparing with $\ve J=\bm{\upsigma}\ve E$ readily yields
\begin{equation}
\boxed{ \bm{\upsigma}^{(nK)} = \nu\, \mathbf N \left( \mathbf I-\nu\mathbf M \right)^{-1} \mathbf G \equiv \nu\,\mathbf Z_c }
\label{eq:sigma_nK_final}
\end{equation}
with
\begin{widetext}
\begin{equation}
Z_{c,ij} = \frac{ N_{in} \left[ \left(1-\nu M_{KK}\right)G_{nj} + \nu M_{nK}G_{Kj} \right] + N_{iK} \left[ \nu M_{Kn}G_{nj} + \left(1-\nu M_{nn}\right)G_{Kj} \right] }{ \left(1-\nu M_{nn}\right) \left(1-\nu M_{KK}\right) - \nu^2M_{nK}M_{Kn}}
\label{eq:Zc_explicit}
\end{equation}
\end{widetext}
with $i,j\,\in \left\{x,y,z\right\}$.
For convenience, we give the explicit integral forms of the matrix elements entering the BGK correction tensor in Methods~\ref{app:explicit}.
For multiple species, the full conductivity tensor is simply the sum of the individual species tensors,
\begin{equation}
\bm{\upsigma} = \sum_s \left( \bm{\upsigma}^{(s,E)} + \bm{\upsigma}^{(s,nK)} \right).
\label{eq:sigma_full}
\end{equation}
To our knowledge, this is the first notation of the full energy and particle conserving BGK collision correction term to the conductivity for drifting and/or anisotropic currents.
Note, that in the isotropic, non-drifting, longitudinal limit, this number- and energy-conserving conductivity reduces to exactly the known scalar expression derived e.g. in \cite{Fried1966, Opher2002}.
\begin{figure*}[tb]
   \centering
   \includegraphics[width=1\linewidth]{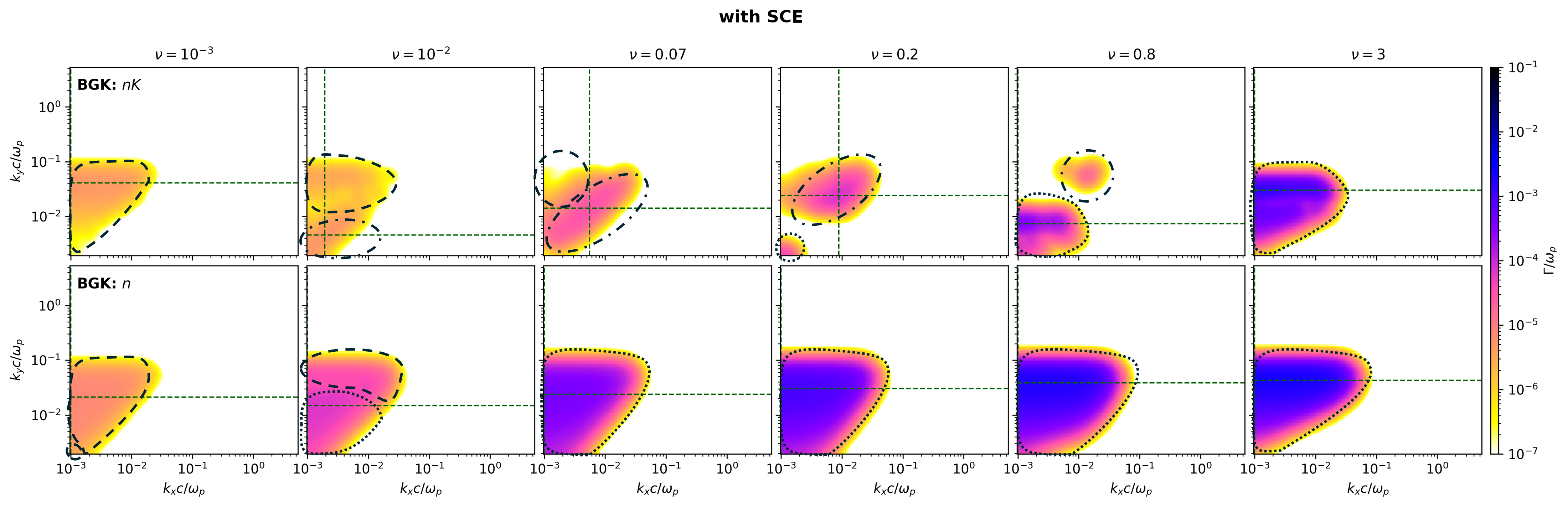}
   \includegraphics[width=1\linewidth]{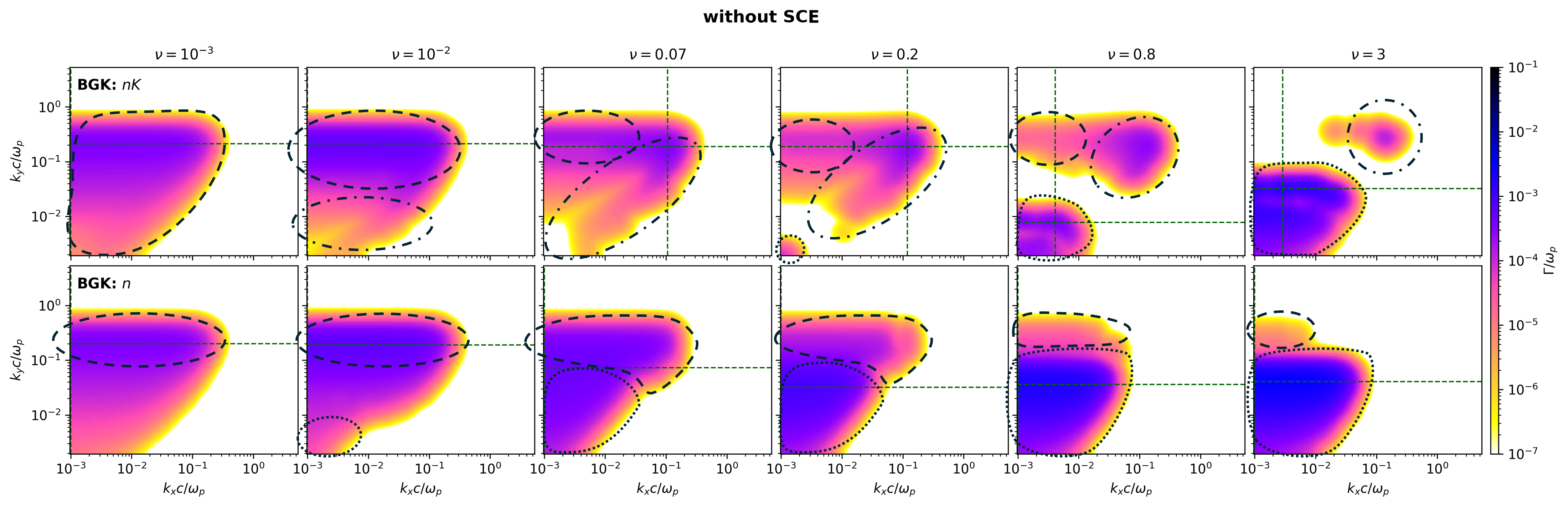}
   \caption{Growth rate maps for the neutral two-current system ($\alpha = 0.02$, $\beta^{(1)}_{th,x}=\beta^{(1)}_{th,y}=0.5$, $\nu^{(1)}=0$, $T^{(2)}=100\unit{eV}$, $\beta^{(1)}_d=0.3$, as a function of $k_x$ and $k_y$, both with (top panel) and without (bottom panel) including the SCE term in Eq.~\ref{eq:dispersion_explicit}.
Dashed crosses mark maximum growth rate position.
Note that for small collision rate in physical systems typically many other physical process may dominate and the simple collision picture becomes unphysical.
They are still shown here and below to visualize the numerically unstable solutions of the dispersion relation.}
   \label{fig:all_2D}
\end{figure*}
\begin{figure}[tb]
   \centering
   \includegraphics[width=1\linewidth]{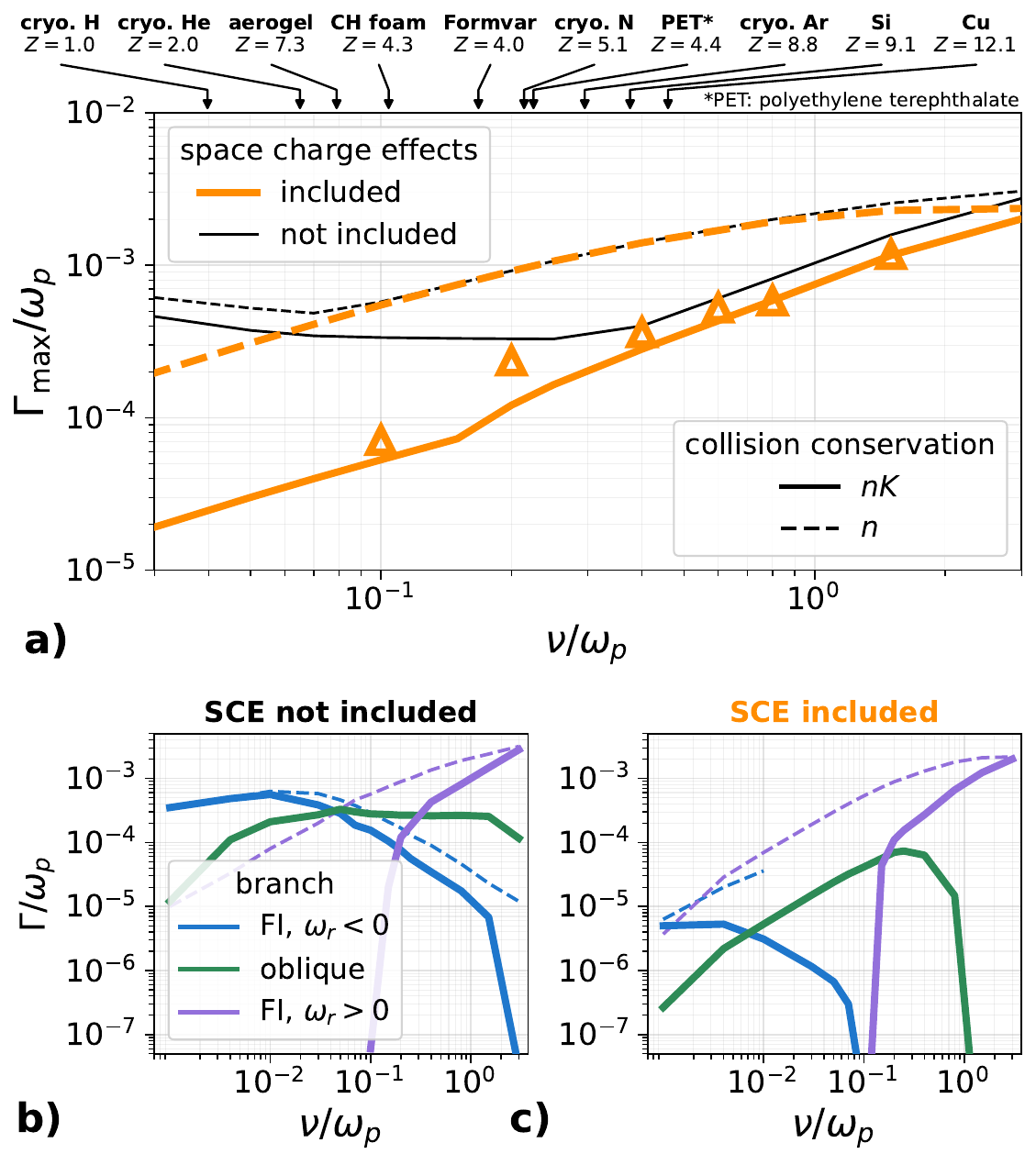}
   \caption{Maximum filamentation growth rates as a function of bulk collision frequency for the density and energy conserving collision operator (solid lines) and density-only conserving one (dashed lines) for the two counter-streaming current example described in the main text.
Panel a) shows maximum growth rate in the full $k$-space within the physically meaningful $\nu$-range, with the calculated collision frequencies at $100\unit{eV}$ marked for some exemplary materials at the top (average ionization states determined as described in Methods from Refs.~\cite{Massacrier2011,Miloshevsky2015,Wu2025} and from an LTE Saha--Boltzmann ionization balance using NIST ionization energies~\cite{Kramida2025}).
Thick orange (thin black) lines show the growth rates including (excluding) SCE.
Triangles show results of particle-in-cell simulations (see Methods).
Panels (b) and (c) show the maximum growth rates separately for the individual branches identified in Fig.~\ref{fig:all_2D} for neglecting SCE (b) and including them (c).}
   \label{fig:all_vertical_profiles_grid}
\end{figure}

The collision correction can physically be interpreted as a self-consistent feedback of the conserved density and laboratory-frame kinetic-energy moments.
The source matrix $\mathbf G$ describes how the electromagnetic field drives density and centered kinetic-energy perturbations, the matrix \(\mathbf M\) describes their collisional self- and cross-coupling, and \(\mathbf N\) converts the resulting moment perturbations back into current.
Restricting the BGK projection to density conservation gives \(Z_{c,ij}^{(n)}=N_{in}G_{nj}/(1-\nu M_{nn})\), recovering the density-conserving result of Ref.~\cite{Hao2009a}.
Energy conservation adds the corresponding energy channel, and its coupling to density through \(M_{nK}\) and \(M_{Kn}\).

We now examine how BGK energy conservation modifies instabilities in a two-current system.
A dilute, fast current $1$ is balanced by a slow bulk return current $2$, both drifting along $x$, with $v_d^{(2)}=-(n^{(1)}/n^{(2)})v_d^{(1)}$.
We treat current $1$ as collisionless and consider only bulk collisions, $\nu_2\equiv\nu$, which cause isotropic heating.
We focus on the temperature-driven regime ($v_d^{(1)} \ll v_{th}^{(1)}$) with isotropic beam temperatures, representative of moderately relativistic laser-plasma interactions \cite{Schoenwaelder2025}.
At ultra-high intensities, beams may instead exhibit narrow longitudinal but broad polarization-direction momentum distributions \cite{Kluge2016b,Kluge2018b}.
Such anisotropic distributions, hot bulk plasmas, and beam-driven regimes are addressed in the Supplemental Material~\cite{SupplementalMaterial}.

For explicit conductivity expressions, we assume drifting anisotropic Maxwellians,
\begin{equation}
f_0^{(s)}(\ve v) = \frac{n_0^{(s)}}{(2\pi)^{3/2}\prod_i v_{th,i}^{(s)}}
\exp\left[-\sum_i \frac{(v_i-\delta_{ix}v_d^{(s)})^2}{2(v_{th,i}^{(s)})^2}\right]. \nonumber
\end{equation}
We restrict $\ve k$ to the $x$--$y$ plane, decoupling in-plane and out-of-plane modes without loss of generality for axisymmetric distributions.
Section~A of the Supplemental Material details this decoupling, while Secs.~C and D derive the conductivity elements in the rotated basis $\hat{\ve e}_\parallel=\ve k/k$, including the density- and energy-conserving BGK corrections~\cite{SupplementalMaterial}.
From Eq.~\eqref{eq:M_def}, the condition for unstable solutions, $\det\mathbf W=0$, yields
\begin{equation}
\left(1+\frac{\ii\sigma_{\parallel\parallel}}{\omega}\right)
\left(1+\frac{\ii\sigma_{\perp\perp}}{\omega}-\frac{k^2}{\omega^2}\right)
+\frac{\sigma_{\parallel\perp}\sigma_{\perp\parallel}}{\omega^2}=0,
\label{eq:dispersion_explicit}
\end{equation}
where the last term represents the space-charge effect (SCE).
For each $\ve k=(k_x,k_y)$, we are left to numerically determine the root $\omega=\omega_r+\ii\Gamma$ with maximum growth rate.

Fig.~\ref{fig:all_2D} shows the numerical results to Eq.~\eqref{eq:dispersion_explicit} for different collision frequencies.
Since the role of SCE is important when discussing energy conservation in collisions, we also show the results for ignoring the SCE term.
At low collision frequencies, transverse filamentation ($k_x=0$) with negative real frequency dominates (dashed lines).
When forward and return currents filament differently, charge separation generates longitudinal electrostatic fields that oppose filamentation \cite{Tzoufras2006b,Hao2009b}, increasingly with collision frequency and more strongly when BGK energy conservation is enforced.
At high collision frequency, transverse filamentation with \textit{positive} real frequency dominates (dotted lines), emerging at small wavenumbers and shifting toward larger growth rates and transverse wavenumbers as collisions increase.
This branch is largely insensitive to SCE, consistent with collision-induced decoupling from electrostatic modes \cite{Hao2009b}, but remains strongly suppressed by energy conservation at moderate collision frequencies.
Together, these effects allow a nearly aperiodic oblique branch to emerge and dominate at intermediate collision frequencies (dash-dotted lines). \\
Fig.~\ref{fig:all_vertical_profiles_grid} shows for each branch separately the maximum growth rate as a function of collision frequency and how these effects influence the global dominant growth rate.
The negative-real-frequency branch is discussed further in the Supplemental Material, which also presents the transverse-wavenumber dependence in Fig.~\ref{fig:nu2_scan_summary_all_cases}~\cite{SupplementalMaterial}.
In contrast to the density conservation alone, the energy conserving BGK term can substantially stabilize the current over a wide range of practically relevant collision frequencies (marked for some exemplary materials).
In fact, at modest collision frequency relevant for most solids, this stabilizing effect seen in panel (c) is dramatically affecting the transverse branches.
Comparison with panel (b) shows that the main effect for the negative frequency branch is an increased SCE induced stabilization.
For the positive frequency branch, the stabilization at small to medium collision frequencies is largely independent on the SCE.
The oblique instability at medium collision frequencies is in the middle: it is suppressed by SCE less than the former, but more than the latter, which leads to a net suppression of instabilities in that range by approximately an order of magnitude compared to density conservation alone.

To validate our model, we performed two-dimensional particle-in-cell simulations over the range of collision frequencies relevant for solids.
The measured growth rates agree with the energy-conserving predictions (Fig.~\ref{fig:all_vertical_profiles_grid}a), falling substantially below the density-only model at intermediate collisionality.
Experimental benchmarks are scarce, as most studies concern collisionless or near-critical-density plasmas, cold beams, or lack quantitative growth-rate comparisons \cite{Bret2009,Gode2017,Scott2017,Romagnani2019,Fiore2010b,Ruyer2020a}.
However, the effect of BGK energy conservation is negligible for very cold or hot beams.
Notably, one recent experiment with laser-accelerated electrons in metal wires showed unexpectedly low growth rates both in x-ray imaging as well as in simulations\cite{Schoenwaelder2025}, consistent with our predictions.

Our results show that enforcing laboratory-frame kinetic-energy conservation can qualitatively alter collisional filamentation by maintaining the coupling between current, density, and energy perturbations. \\ In the counterstreaming system considered here, this stabilizes the currents over a broader range of collision frequencies, and suppresses transverse negative and positive frequency filamentation, which allows a nearly aperiodic oblique mode to emerge and dominate.
Both the maximum growth rate and the characteristic filament scale therefore differ substantially from predictions based on density conservation alone.
Generally, the role of energy conservation is complex and depends on the current properties.
As shown in the Supplemental Material~\cite{SupplementalMaterial}, there are specific cases in which it can even increase the growth rate. \\ The corresponding conductivity correction, Eq.~\eqref{eq:sigma_nK_final}, provides a compact kinetic description of this coupling for general equilibrium distributions.
It allows for closed-form results, e.g. for non-relativistic drifting anisotropic Maxwellians and a scalar collision frequency as presented here; extensions to more complete collision operators that would consider also inter-species energy transfer and collisions with other species than heavy ions will generally require numerical evaluation of the underlying kinetic integrals.

\section*{Acknowledgements}
This work has been partially supported by the EUROfusion Enabling Research Project No. CfP-FSD-AWP26-ENR-01 “Conceptual design for a European High Power Laser Fusion Research Facility” ( HiPER+RF), funded by the European Union via the Euratom Research and Training Programme (Grant Agreement No. 101052200—EUROfusion).

\section*{Code availability}

For the particle-in-cell simulations we used the code PIConGPU.
It is open source and available online \cite{Unknown2017b}.

\renewcommand{\bibsection}{\section*{References}}
\bibliographystyle{apsrev4-1}
\bibliography{references}

\setcounter{equation}{0}
\renewcommand{\theequation}{M\arabic{equation}}

\renewcommand{\theHequation}{M\arabic{equation}}
\section*{Methods}

\subsection{Construction of the particle- and energy-conserving BGK operator}
\label{app:conservation}

Throughout the Methods we suppress species indices for brevity.

Here we construct the BGK operator used in the main text from the requirements of particle-number and laboratory-frame kinetic-energy conservation.
For a single species, we assume a BGK relaxation toward the minimal subspace spanned by the two conserved moments, particle number and kinetic energy.
This restricts the relaxed perturbation to a linear combination of the corresponding momentum-space basis functions.
Higher-order functions of the particle energy could in principle be added while preserving the same two moments, but would introduce additional kinetic degrees of freedom not fixed by these conservation laws and are therefore not included in the present two-moment BGK closure.

For a single species, the relativistic laboratory-frame single-particle specific kinetic energy is
\begin{equation}
H(\ve p) = \gamma(\ve p)-1 .
\label{eq:H_rel_lab_def_s}
\end{equation}

We use the centered energy variable \(K(\ve p)\) defined in Eq.~\eqref{eq:K_main_def}, with \(\bar H_{\mathrm{lab}}\equiv\langle H\rangle_0\).
By construction,
\begin{equation}
\int K(\ve p)f_0(\ve p)\,\dd^3p = 0 .
\label{eq:K_rel_centered_property_s}
\end{equation}
The corresponding equilibrium variance is
\begin{equation}
\mathcal K_{2} = \frac{1}{n_0} \int \left(K(\ve p)\right)^2 f_0(\ve p)\,\dd^3p .
\label{eq:E2_rel_lab_def_s}
\end{equation}

The two conserved perturbation moments are those collected in \(\delta\ve m\) in Eq.~\eqref{eq:nK_def}.
In the following we write \(\delta K\equiv\delta\mathcal K\) for the centered kinetic-energy moment.
Note, that \(K(\ve p)\) is the centered single-particle energy coordinate, whereas \(\delta K\) is the corresponding macroscopic energy moment of the perturbation.

A BGK operator relaxes \(\delta f\) toward a projected perturbation while leaving the selected conserved moments unchanged.
We start from the ansatz
\begin{equation}
\mathcal{C}[\delta f] = \nu \left[ \delta f - f_0 \left( \eta_n\delta n + \eta_KK\delta K \right) \right],
\label{eq:bgk_general_nK}
\end{equation}
where the coefficients \(\eta_n\) and \(\eta_K\) are determined by the conservation requirements.

The particle-number component of the conservation condition in Eq.~\eqref{eq:moment_basis} gives
\begin{align}
0 & = \nu \Biggl[ \delta n - \eta_n\delta n \int f_0\,\dd^3p \nonumber \\
  & \qquad - \eta_K\delta K \int K(\ve p)f_0(\ve p)\,\dd^3p \Biggr].
\end{align}
Using the normalization of \(f_0\) and Eq.~\eqref{eq:K_rel_centered_property_s}, this fixes
\begin{equation}
\eta_n = \frac{1}{n_0} .
\label{eq:bgk_eta_n_coefficient}
\end{equation}

The energy component of Eq.~\eqref{eq:moment_basis} similarly gives, upon substitution of Eq.~\eqref{eq:bgk_general_nK},
\begin{align}
0 & = \nu \Biggl[ \delta K - \eta_n\delta n \int K(\ve p)f_0(\ve p)\,\dd^3p \nonumber \\
  & \qquad - \eta_K\delta K \int \left(K\right)^2 f_0\,\dd^3p \Biggr].
\end{align}
The first projected contribution vanishes because \(K\) is centered, and from Eq.~\eqref{eq:E2_rel_lab_def_s},
\begin{equation}
\int \left(K\right)^2 f_0\,\dd^3p = n_0 \mathcal K_{2} .
\label{eq:E2_rel_identity_s}
\end{equation}
Hence
\begin{equation}
\eta_K = \frac{1}{ n_0 \mathcal K_{2} }.
\label{eq:bgk_eta_K_coefficient}
\end{equation}

The minimal particle- and energy-conserving BGK operator is therefore
\begin{equation}
\mathcal{C}[\delta f] = \nu \left[ \delta f - f_0 \left( \frac{\delta n}{n_0} + \frac{ K\delta K }{ n_0\mathcal K_{2} } \right) \right].
\label{eq:bgk_rel_derived}
\end{equation}
Using \(\ve\psi\) from Eq.~\eqref{eq:psi_def} and \(\mathbf S\) from Eq.~\eqref{eq:moment_gram_matrix}, for which centering gives \(\mathbf S^{-1}=\operatorname{diag}(1,1/\mathcal K_2)\), Eq.~\eqref{eq:bgk_rel_derived} is identical to Eq.~\eqref{eq:bgk_nK}.

Finally, since
\begin{equation}
H = K + \bar H_{\mathrm{lab}},
\label{eq:Hrel_Krel_eps_s}
\end{equation}
conservation of particle number and of the centered-energy moment also implies conservation of the full laboratory-frame relativistic kinetic-energy moment:
\begin{equation}
\begin{aligned}
\int H C[\delta f]\,\dd^3p & = \int K \mathcal{C}[\delta f]\,\dd^3p \\
                          & \quad + \bar H_{\mathrm{lab}} \int \mathcal{C}[\delta f]\,\dd^3p = 0 .
\end{aligned}
\label{eq:bgk_rel_H_conservation}
\end{equation}
Thus Eq.~\eqref{eq:bgk_rel_derived} is the minimal BGK relaxation operator that preserves, for the single-species response, both particle number and laboratory-frame relativistic kinetic energy.

\subsection{Explicit component forms}
\label{app:explicit}

The individual quantities entering Eq.~\eqref{eq:Zc_explicit} can be written as scalar component integrals.
The components of the response matrix \(\mathbf N\) are
\begin{align}
N_{in} & = \frac{q}{n_0} \int v_i \frac{f_0}{D} \,\dd^3p, \\
N_{iK} & = \frac{q}{n_0\mathcal K_2} \int v_i K \frac{f_0}{D} \,\dd^3p.
\end{align}
The four components of the moment-feedback matrix are
\begin{align}
M_{nn} & = \frac{1}{n_0} \int \frac{f_0}{D} \,\dd^3p, \\
M_{nK} & = \frac{1}{n_0\mathcal K_2} \int K \frac{f_0}{D} \,\dd^3p, \\
M_{Kn} & = \frac{1}{n_0} \int K \frac{f_0}{D} \,\dd^3p, \\
M_{KK} & = \frac{1}{n_0\mathcal K_2} \int K^2 \frac{f_0}{D} \,\dd^3p.
\end{align}
Finally, since
\begin{equation}
F_{\ell j} = \left( 1-\frac{1}{\omega} \sum_r k_r v_r \right)\delta_{\ell j} + \frac{k_\ell v_j}{\omega},
\end{equation}
from Eq.~\eqref{eq:F_def}, the two field-to-moment source components can be written as
\begin{widetext}
   \begin{align}
       G_{nj} & = -q \int \frac{1}{D} \left[ \left( 1-\frac{1}{\omega}\sum_r k_r v_r \right) \frac{\partial f_0}{\partial p_j} + \frac{v_j}{\omega} \sum_\ell k_\ell \frac{\partial f_0}{\partial p_\ell} \right] \,\dd^3p, \\
       G_{Kj} & = -q \int \frac{K}{D} \left[ \left( 1-\frac{1}{\omega}\sum_r k_r v_r \right) \frac{\partial f_0}{\partial p_j} + \frac{v_j}{\omega} \sum_\ell k_\ell \frac{\partial f_0}{\partial p_\ell} \right] \,\dd^3p,
   \end{align}
\end{widetext}
where \(i,j,\ell,r\in\{x,y,z\}\) and \(D\) is the resonance denominator defined in the main text immediately before Eq.~\eqref{eq:delta_f}.

\subsection{PIC simulation setup}

The particle-in-cell simulations were initialized with two almost homogeneous counter-streaming currents on a grid of $640\times 6400$ cells.
The forward current density was set to $8\,n_c$, the density of the return current was set to $400\,n_c$ with some arbitrary density normalization $n_c$.
An ion species of $408\,n_c$ was also added.
We used 100 randomly positioned particles per cell for the return current and ions species, and 20 particles per cell for the forward current.
The cell size was set to 16 cells per plasma period $\omega_p=\sqrt{n_c}$, the timestep was set to $\sqrt{.5}/16$ of the plasma period to satisfy the Courant condition.
The collision rate was controlled by setting a respective fixed value for the coulomb logarithm.
The random initialization of particle positions provides for a seed of all spatial frequencies.
Subsequently, the dominant filamentation wavenumber will grow fastest and hence is determining the observed growth rate.
For the smallest collision frequency (i.e. smallest growth rate), we used 10 times the particle number per cell, in order to suppress Poisson noise.

\subsection{Ionization states and material collision frequencies}

The materials in Fig.~\ref{fig:all_vertical_profiles_grid} were evaluated at an electron temperature of $T_e=100\unit{eV}$ and at their respective material ion densities.

For carbon and oxygen in solid-density polymers, we used the dense-plasma ionization model of Massacrier \textit{et al.}~\cite{Massacrier2011}.
Their model accounts for pressure ionization, Coulomb interactions, and electron degeneracy.
Interpolation of the charge-state populations around $T_e=100\unit{eV}$ and densities of order $1\unit{g\,cm^{-3}}$ gives approximately $\bar Z_{\rm C}=4.2$ and $\bar Z_{\rm O}=5.8$.
These values were used for the carbon- and oxygen-containing solid-density polymer targets.

For solid-density silicon, we used $\bar Z_{\rm Si}=9.1$, obtained by Wu \textit{et al.}~\cite{Wu2025}.
In that work, the mean ionization is calculated from detailed-configuration Saha--Boltzmann equations, with ionization potentials modified by the ionization-potential depression.

For solid-density copper, we used $\bar Z_{\rm Cu}\simeq12.1$, based on the collisional-radiative calculations of Miloshevsky and Hassanein~\cite{Miloshevsky2015}, which include the density-induced lowering of bound states.
At $T_e\simeq100\unit{eV}$ and the solid density of copper, the charge-state distribution is centered around approximately twelvefold-ionized Cu.

For the cryogenic materials H, He, N, and Ar, the ionization balance was evaluated in local thermodynamic equilibrium (LTE) by solving the successive Saha--Boltzmann equations at $T_e=100\unit{eV}$ and the corresponding ion number densities.
The isolated-ion ionization energies were taken from the NIST Atomic Spectra Database~\cite{Kramida2025}.
The resulting mean charge states used in the material comparison were $\bar Z_{\rm H}=1$, $\bar Z_{\rm He}=2$, $\bar Z_{\rm N}=5.1$, and $\bar Z_{\rm Ar}=8.8$.

In addition to solid-density and cryogenic targets, we considered two representative low-density materials commonly employed in high-energy-density experiments: silica aerogel and plastic foam.
For silica aerogel (SiO$_2$), we assumed a mass density of $\rho=20\unit{mg\,cm^{-3}}$, representative of low-density aerogel targets~\cite{Patterson2018}.
At $T_e=100\unit{eV}$, we adopted the approximate charge states $\bar Z_{\rm Si}=9$ and $\bar Z_{\rm O}=6$, giving a mean charge $\bar Z_{\rm mix}=7.0$ and an effective collision charge $Z_{\rm eff}=7.29$.
For the plastic foam, we assumed a CH composition with $\rho=100\unit{mg\,cm^{-3}}$, representative of foam targets used in inertial-fusion studies~\cite{Streit2003}.
We adopted $\bar Z_{\rm C}=5$ and $\bar Z_{\rm H}=1$, yielding $\bar Z_{\rm mix}=3.0$ and $Z_{\rm eff}=4.33$.
The charge states of these low-density materials are representative estimates, rather than results of dedicated ionization-balance calculations at the specified densities.

For compounds, the mean number of free electrons per nucleus was constructed from the elemental values according to the stoichiometric composition,
\begin{equation}
    \bar Z_{\rm mix}
    =
    \frac{\sum_s N_s \bar Z_s}{\sum_s N_s},
\end{equation}
where $N_s$ is the number of nuclei of constituent $s$.
We used CH$_2$ for polyethylene, C$_5$H$_8$O$_2$ for Formvar, C$_{10}$H$_8$O$_4$ for PET, C$_{22}$H$_{10}$N$_2$O$_5$ for Kapton, SiO$_2$ for silica aerogel, and CH for the plastic foam.
The total ion number density was calculated from the mass density $\rho$ as
\begin{equation}
    n_i
    =
    \frac{\rho N_A \sum_s N_s}
         {\sum_s N_s M_s},
\end{equation}
where $N_A$ is the Avogadro constant and $M_s$ is the molar mass of constituent $s$.
For the silica aerogel and CH foam, this gives $n_i=6.01\times10^{20}\unit{cm^{-3}}$ and $n_i=9.25\times10^{21}\unit{cm^{-3}}$, respectively.

For evaluating the electron-ion collision frequency, the first and second moments of the charge distribution were distinguished.
The electron density is determined by the mean charge, $n_e=n_i\bar Z$, whereas the $Z^2$ weighting of electron-ion scattering is represented by
\begin{equation}
    Z_{\rm eff}
    =
    \frac{\langle Z^2\rangle}{\langle Z\rangle}.
\end{equation}
For mixtures,
\begin{equation}
    Z_{\rm eff}
    =
    \frac{\sum_s N_s \langle Z_s^2\rangle}
         {\sum_s N_s \langle Z_s\rangle},
\end{equation}
so that $n_i\bar Z Z_{\rm eff}$ reproduces the $\sum_s n_s\langle Z_s^2\rangle$ weighting entering the electron-ion collision rate.
In the numerical material estimates, the charge-state distributions were approximated by their elemental mean charges, $\langle Z_s^2\rangle\simeq\bar Z_s^2$, neglecting fluctuations of the ionic charge state.
The resulting collision frequencies were evaluated using the Coulomb-logarithm\cite{Gericke2002}
\begin{equation}
\ln\Lambda
=
\frac12\ln\left[
1+
\frac{T_e/n_e+(3\alpha\hbar/n_i)^{2/3}}
     {(Z_{\rm eff}\alpha\hbar/2T_e)^2+\hbar^2/(8T_e)}
\right],
\end{equation}
where $\alpha$ is the fine-structure constant and $\hbar$ is the normalized reduced Planck constant.
We chose this, because the classical Spitzer Coulomb logarithm can approach unity or become unphysical.
The expression above accounts for the finite ion-sphere radius, yielding a smooth, positive Coulomb logarithm without imposing an arbitrary lower cutoff.

Note, that the resulting collision frequency estimates are intended as a physically relevant collision-frequency approximation only.
In particular, the ionization state and collision frequency of porous, low-density materials depend on their actual thermodynamic state, and the assumed values do not account for spatial inhomogeneities or uncertainties in the ionization balance.

\subsection{Limits}

The general kinetic and electromagnetic formulation used in this work is linear, homogeneous, unmagnetized, and three-dimensional in wave-vector and momentum space.
The closed-form expressions and numerical examples presented here additionally assume to non-relativistic drifting anisotropic Maxwellians, and a scalar collision frequency.
Zero-th order relativistic mass increase of the electrons can be easily included by replacing the mass in the plasma frequency and the definition of $n_\mathrm{eff}$ by a constant relativistic mass $\gamma^{(s)}m^{(s)}$.
However, for ultra-relativistic currents the integrals entering \(\mathbf N\), \(\mathbf M\), and \(\mathbf G\) generally no longer reduce to simple closed-form expressions and need to be evaluated numerically.
Yet, it is important to note that even for relativistic laser intensity the fast electron current of a diverging electron beam in simulations still shows as non-relativistic Maxwell distribution up to close to the speed of light, as can be constructed also analytically from~\cite{Kluge2018c}.

The present treatment does not describe nonlinear saturation, filament coalescence, finite target geometry, evolving background heating, velocity-dependent Coulomb scattering, or collisions between multiple species.
The present BGK projector conserves the laboratory-frame kinetic energy within a scalar-\(\nu\) BGK closure but does not take into account angular diffusion in velocity space, energy diffusion, and momentum-conserving channels, relevant e.g. for collisions between species with comparable mass.
When collisions are not elastic, energy and momentum must be conserved not only for light particles but for the full set of particles, which greatly increases complexity of the collision operator and makes an analytical formulation for the conductivity tensor impossible.
\end{document}